\newif\ifmulticol	\multicoltrue
\newif\ifshowgit	\showgittrue		
\newif\ifgitlocal	\gitlocaltrue		
\newif\ifbiblatex	\biblatextrue		
\newif\ifbibnum		\bibnumtrue			
\newif\iflineno		\linenofalse
\newif\iftoc		\tocfalse

\newif\iflucida		\lucidafalse
\newif\ifcm			\cmfalse
\newif\iflibertine	\libertinetrue
\newif\ifcharter	\charterfalse

\newcommand*{\secnumdepth}{0}			
\newcommand*{\tocdepth}{2}				
\newcommand*{\reftitle}{References}		
\newcommand*{\mytitleskip}{-0.2in}		

\newcommand*{\mydocfontsize}{\ifcharter11pt\else\iflibertine11pt\else10pt\fi\fi}
\newcommand*{\setcol}{\ifmulticol twocolumn\else onecolumn\fi}

\documentclass[\mydocfontsize,\setcol]{article}

\newcommand*{\pdfauthor}{Steven A.~Frank}
\newcommand*{\pdftitle}{Automatic differentiation and the optimization of differential equation models in biology}

\usepackage{tabularx}
\usepackage{booktabs}
\usepackage{multirow}
\usepackage[percent]{overpic}

\input adiff.sty

\newcommand*{\bx}{\bmr{x}}
\newcommand*{\bxt}{\bx_t}
\newcommand*{\bX}{\bmr{X}}
\newcommand*{\bp}{{\bmr{p}}}
\newcommand*{\gL}{\nabla L}
\newcommand*{\ggL}{\nabla^2 L}
\newcommand*{\bu}{\bmr{u}}
\newcommand*{\bS}{\bmr{S}}
\newcommand*{\bb}{\bmr{b}}

\usepackage{bm}
\newcommand{\bmr}[1]{\bm{\mathrm{#1}}}

\DeclarePairedDelimiter\abs{\lvert}{\rvert}
\DeclarePairedDelimiter\norm{\lVert}{\rVert}
\DeclarePairedDelimiter\angb{\langle}{\rangle}
\DeclarePairedDelimiter\lrb{\lbrack}{\rbrack}
\DeclarePairedDelimiter\lr{\lparen}{\rparen}
\DeclarePairedDelimiter\lrbr{\lbrace}{\rbrace}

\makeatletter
\let\oldabs\abs \def\abs{\@ifstar{\oldabs}{\oldabs*}}
\let\oldnorm\norm \def\norm{\@ifstar{\oldnorm}{\oldnorm*}}
\let\oldangb\angb \def\angb{\@ifstar{\oldangb}{\oldangb*}}
\let\oldlrb\lrb \def\lrb{\@ifstar{\oldlrb}{\oldlrb*}}
\let\oldlr\lr \def\lr{\@ifstar{\oldlr}{\oldlr*}}
\let\oldlrbr\lrbr \def\lrbr{\@ifstar{\oldlrbr}{\oldlrbr*}}
\makeatother

\newcommand*{\dd}{\textrm{d}}

\newcommand*{\EEq}[1]{Eqn~\ref{eq:#1}}
\newcommand*{\Eq}[1]{eqn~\ref{eq:#1}}

\newcommand*{\dovr}[2]{\frac{\dd #1}{\dd #2}}

\newcommand*{\prt}{\partial}
\newcommand*{\povr}[2]{\frac{\prt #1}{\prt #2}}

\newcommand*{\Figure}[1]{Figure~\ref{fig:#1}}
\newcommand*{\Fig}[1]{Fig.~\ref{fig:#1}}

\newcommand*{\boldrule}{\hrule height 1.2pt}
\newcommand*{\noterule}{\medskip\boldrule\medskip}	

\usepackage{tikz}
\usetikzlibrary{decorations.pathmorphing}
\usetikzlibrary{arrows}
\usetikzlibrary{positioning}

\newcommand*{\mytitle}{\pdftitle}

\newcommand*{\myauthor}{Steven A.\ Frank}
\newcommand*{\myaddress}{Department of Ecology and Evolutionary Biology, University of California, Irvine, CA 92697--2525, USA}
\newcommand*{\mycontact}{\href{https://stevefrank.org}{https://stevefrank.org}}

\newcommand*{\mykeywords}{Automatic differentiation, optimization, fitness landscape, statistical inference, differential equation modeling, ecological dynamics, evolutionary dynamics}

\setabstract{-0.07}{\iftoc-2.0\else0.5\fi}{%
A computational revolution unleashed the power of artificial neural networks. At the heart of that revolution is automatic differentiation, which calculates the derivative of a performance measure relative to a large number of parameters. Differentiation enhances the discovery of improved performance in large models, an achievement that was previously difficult or impossible. Recently, a second computational advance optimizes the temporal trajectories traced by differential equations. Optimization requires differentiating a measure of performance over a trajectory, such as the closeness of tracking the environment, with respect to the parameters of the differential equations. Because model trajectories are usually calculated numerically by multistep algorithms, such as Runge-Kutta, the automatic differentiation must be passed through the numerical algorithm. This article explains how such automatic differentiation of trajectories is achieved. It also discusses why such computational breakthroughs are likely to advance theoretical and statistical studies of biological problems, in which one can consider variables as dynamic paths over time and space. Many common problems arise between improving success in computational learning models over performance landscapes, improving evolutionary fitness over adaptive landscapes, and improving statistical fits to data over information landscapes.
}

\begin{document}

\mymaketitle

\iftoc\mytoc{-24pt}{\newpage}\fi

\section{Introduction}

Theoretical studies often analyze improvement on a performance surface. What phenotypes enhance fitness? What differential equation model best describes ecological or biochemical dynamics? What interventions improve the health of organisms or ecosystems?

Similarly, inductive methods of statistics and machine learning optimize prediction, minimize error, or maximize the use of information \autocite{mcelreath15statistical,goodfellow16deep}. Natural selection can itself be thought of as nature's inductive process of improving the fit of organisms to their environment \autocite{frank09natural,frank20the-inductive}.

Most of these problems reduce to finding a path that enhances performance \autocite{08encyclopedia}. In practice, that often means changing model parameter values in the direction along which the slope of performance tilts most strongly toward improvement. Ideally, that best slope is found by differentiating the performance surface with respect to the parameters of the model or the traits of the organism.

Taking the derivative of a performance function with respect to its parameters is, in principle, a simple process. Performance functions typically arise by composition of basic mathematical operations such as addition, multiplication, and exponentiation. Standard procedures yield the derivative for each basic operation. The overall derivative follows by the chain rule for differentiation.

Two aspects hinder the calculation of derivatives. First, performance measures often follow from long chains of operations embedded in complex algorithms. For example, performance may depend on a measure of the distance between some optimal trajectory and the trajectory of a system of differential equations specified by a model with many parameters.

In practice, one often calculates the model trajectory for a given parameter set by a numerical algorithm. The derivative of the distance measure between the optimal and realized trajectories must be calculated through the operations of the numerical algorithm, which is typically embedded in the code of a computer program. Efficiently updating the parameters to reduce the distance of the model's trajectory from the optimal trajectory benefits from an automatic method to calculate the derivative of that distance with respect to the parameters \autocite{griewank08evaluating,baydin18automatic,margossian19a-review}.

The second difficulty for calculation arises because performance functions often depend on large numbers of parameters. For example, the modern artificial intelligence revolution arose in part from expanding neural networks to multiple deeply connected layers with millions of parameters. It only became possible to calculate an improving pathway of parameter values with sufficient speed after the development of efficient methods for automatic differentiation of performance functions embedded within the calculations of complex computer code \autocite{goodfellow16deep}.

Conceptually, automatic differentiation provides insight into the topography of performance surfaces. In biology, many problems of evolutionary dynamics turn on the adaptive topography that maps phenotypes to fitness \autocite{stadler02fitness,malan21a-survey}. The close similarity of evolutionary dynamics and the many other problems that depend on the topography of performance surfaces provides insight into how natural selection designs organisms and into the prospects and limits of optimization models in biology. In other words, automatic differentiation will help us to learn more about the shape of optimization problems in both natural history and in models of evolutionary biology.

This article focuses on the benefits of automatic differentiation for differential equation modeling and for broader problems in theoretical biology. Highlights include the opportunities for new kinds of theoretical approaches, a better understanding of current developments in statistics and machine learning, and some possible directions for future research motivated by improved understanding of multidimensional performance surfaces.

I start with an overview of differentiation in optimization. I next turn to the techniques and advantages of automatic differentiation for optimizing differential equation models \autocite{chen18neural,rackauckas20universal}. I then present two examples to illustrate how the optimization of differential equations by automatic differentiation may enhance the future study of various topics in biology.

The first example fits various differential equation models to the classic data on predator-prey population cycles of lynx and hare. The second example searches for transcription factor network designs that solve the challenge of maintaining an internal circadian rhythm. The internal rhythm must buffer against the intrinsic stochasticity of biochemical dynamics, entrain to an erratic external circadian signal when available, and maintain the internal rhythm when the external signal is absent.

Slopes and curvatures calculated by differentiation also provide the basis for enhanced sensitivity analysis \autocite{mester22differential}. In biology, sensitivity plays a key role in understanding robustness, variability, and evolutionary dynamics. In inference, sensitivity influences the degree of belief in conclusions drawn from models and data, often analyzed by Bayesian methods.

\section{Differentiation in optimization}

Optimizing a function $L$ with respect to a parameter vector $\bp$ sets a classic optimization problem. For example, what parameters for a system of differential equations minimize the distance between the system's trajectory and the observed trajectory of fluctuating lynx and hare populations? In this case, the total distance, $L$, might be the sum of the squared deviations between system's trajectory and the observed trajectory when measured at series of temporal intervals.

In general, it is relatively easy to find a local optimum near some initial point in parameter space and relatively difficult to find a global optimum over the full space of potential parameter values. Many optimization techniques exist. The best methods for a particular problem depend on the shape of the performance surface \autocite{08encyclopedia,reddi19on-the-convergence,ruder17an-overview}.

For problems with continuous performance surfaces, using the derivative of $L$ with respect to the parameters, $\bp$, often greatly enhances the search for better parameters. The vector of partial derivatives of $L$ with respect to the parameters is the gradient or slope of the performance surface, $\gL$.

The gradient describes the change in performance with respect to the change in parameters. Thus, the gradient provides much information about how to update parameter values in an iterative search for improving performance. Knowing the gradient does not by itself solve the problem of getting stuck in a local optimum. However, fast calculation of the gradient often transforms problems from hard and beyond reasonable study into ones that can be analyzed relatively easily \autocite{goodfellow16deep,reddi19on-the-convergence,ruder17an-overview}.

Fast calculation of derivatives by automatic methods may also allow calculation of the second derivatives, which describe the curvature of the performance surface. For a problem with $n$ parameters and a single-valued performance measure, $L$, we write $\ggL$ for the $n\times n$ Hessian matrix of partial second derivatives. The Hessian matrix provides insight into many aspects of optimization.

\begin{figure*}[t]
\centering
\includegraphics[width=6in]{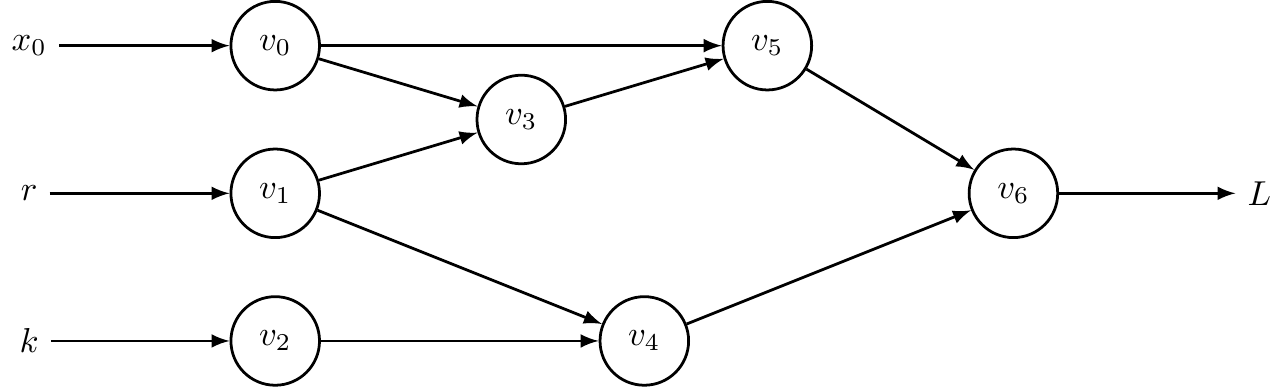}
\caption{Computational graph to calculate $L(x_h)=(1-x_h)^2$ for \Eq{euler} with $t=0$ and $h=0.01$, in which optimization seeks to minimize the loss function, $L$. Table 1 shows the intermediate results, $v_i$, with $v_6=x_h$. This computational graph can be recursively expanded $n$ times by using $v_6=x_t$ as the starting point for additional applications of \Eq{euler} to calculate $x_{nh}$ for any integer, $n$. The resulting graph quickly becomes too large to visualize or to trace the pathways of calculation without some automatic method. However, computer code readily expands such graphs and calculations.}
\label{fig:graph}
\end{figure*}

\section{Optimizing differential equations}

Differential equations provide the primary modeling approach for many problems \autocite{edelstein--keshet88mathematical,ellner06dynamic}. Organismal traits often follow trajectories over time as individuals develop and respond to the environment. Systems of gene regulation, biochemistry, and physiology describe temporal changes in molecular concentrations and biological outputs. Continuous models of population genetics describe temporal changes in allelic and genotypic frequencies. Ecological dynamics describe temporal changes in populations and interacting species.

The differential equations describe a vector of values, $\bx$, that depend on time, $t$, on a parameter vector, $\bp$, on initial conditions, and on various other inputs. If we write the system values at time $t$ as $\bxt$, then we can abbreviate the differential equations as
\begin{equation*}
  \bxt' = f(\bxt,\bp),
\end{equation*}
in which the prime denotes differentiation with respect to $t$.

Many studies seek a parameter vector, $\bp$, that optimizes a temporal trajectory of values, $\bxt$, over some time span. Analysis optimizes a performance function, $L\lr{\bX}$ with respect to $\bp$ for the system values at various times, $\bX=\lrbr{\bx_{t_1},\bx_{t_2},\ldots}$.

This section illustrates automatic differentiation for the simple differential equation
\begin{equation}\label{eq:diffeq}
  x'_t = f(x_t,\bp)=r\lr{k-x_t}
\end{equation}
with parameter vector $\bp=\lrbr{r,k}$. We use the performance function $L(x_T)=(x_T-c)^2$, the squared Euclidean distance between the system's value at particular time, $T$, and the target value, $c$, with the goal of minimizing the performance function.

\EEq{diffeq} has a simple explicit solution, which we could use to find optimum values of $\bp$. However, in most applications, the system of differential equations must be evaluated numerically. Optimization must be done by choosing some parameters, numerically calculating the solution, $x_t$, and the loss function, $L$, and then updating the parameters to improve performance \autocite{chen18neural,rackauckas20universal}. The optimization procedure consists of repeated rounds of calculation and parameter updating. The goal is to improve performance and, ideally, to find an optimal parameter vector that minimizes the loss function.

To illustrate the typical optimization cycle, we evaluate \Eq{diffeq} numerically by Euler's method
\begin{equation}\label{eq:euler}
  x_{t+h} \approx x_t + hf(x_t,\bp) = x_t + hr\lr{k-x_t},
\end{equation}
in which $h$ is the stepsize over which we update values. In practice, one uses better numerical methods to approximate the trajectory that $x_t$ follows over time \autocite{leveque07finite}. However, the concepts by which one applies automatic differentiation do not depend on the numerical technique, so Euler's method is sufficient to illustrate how one uses automatic differentiation to aid in optimizing the parameters of differential equations.

\begin{table*}
  \centering
  \renewcommand{\arraystretch}{1.2}
  \caption{Calculations through the computational graph in \Fig{graph}. The left box traces the calculation of values forward from the inputs to the output, yielding the final value for $L$. The center box traces the forward chain rule calculation of the derivatives with respect to the input parameter, $r$, yielding $\dot{L}=\prt{L}/\prt{r}$ at the bottom. The variable $z$ represents the left side of each equation. A separate derivative trace is required for each input parameter. The right box traces the reverse chain rule calculation of the derivatives, starting from the bottom and proceeding to the top. Reverse mode has the advantage that a single trace calculates the partial derivatives of $L$ with respect to all of the input parameters. A constant value $h=0.01$ is used in calculations. Structure of table based on Baydin et al.'s (\citeyear{baydin18automatic}) Table 2.}
  \label{tableTrace}
  \begin{minipage}[c]{0.27\textwidth}
    {\footnotesize
    \begin{tabularx}{\textwidth}{p{0.2mm}p{0.2mm}p{15mm}X}
      \toprule
      \multicolumn{4}{l}{Forward value trace}\\
      \multirow{9}{2mm}{\begin{tikzpicture}\draw[->,>=triangle 60,thick](0,0)--(0,-3.8);\end{tikzpicture}}
      & $v_0$ & $=x_0$ & $=0$\\
      & $v_1$ & $=r$ & $=0.1$\\
      & $v_2$ & $=k$ & $=1$\\
      \cmidrule{2-4}
      & $v_3$ & $=hv_0v_1$ & $=0$\\
      & $v_4$ & $=hv_1v_2$ & $=0.001$\\
      & $v_5$ & $=v_0-v_3$ & $=0$\\
      & $v_6$ & $=v_4+v_5$ & $=0.001$\\
      \cmidrule{2-4}
      & $L$ & $=(1-v_6)^2$ & $=0.998001$\\
      \bottomrule
    \end{tabularx}}
  \end{minipage}
  \begin{minipage}[c]{0.33\textwidth}
    \setlength{\fboxsep}{0pt}\colorbox{gray!20}
    {\footnotesize
    \begin{tabularx}{\textwidth}{p{0.2mm}p{0.2mm}p{25mm}X}
      \toprule
      \multicolumn{4}{l}{Forward derivative trace $(\dot{z}=\prt z/\prt r)$}\\
      \multirow{9}{2mm}{\begin{tikzpicture}\draw[->,>=triangle 60,thick](0,0)--(0,-3.8);\end{tikzpicture}} 
      & $\dot{v}_0$ & $=\dot{x}_0$ & $=0$\\
      & $\dot{v}_1$ & $=\dot{r}$ & $=1$\\
      & $\dot{v}_2$ & $=\dot{k}$ & $=0$\\
      \cmidrule{2-4}
      & $\dot{v}_3$ & $=h\lr{v_0\dot{v}_1+\dot{v}_0v_1}$ & $=0$\\
      & $\dot{v}_4$ & $=h\lr{v_1\dot{v}_2+\dot{v}_1v_2}$ & $=0.01$\\
      & $\dot{v}_5$ & $=\dot{v}_0-\dot{v}_3$ & $=0$\\
      & $\dot{v}_6$ & $=\dot{v}_4+\dot{v}_5$ & $=0.01$\\
      \cmidrule{2-4}
      & $\dot{L}$ & $=-2\lr{1-v_6}\dot{v}_6$ & $=-0.01998$\\
      \bottomrule
      \end{tabularx}}
  \end{minipage}
  \begin{minipage}[c]{0.38\textwidth}
    \setlength{\fboxsep}{0pt}\colorbox{gray!20}
    {\footnotesize
    \begin{tabularx}{\textwidth}{p{0.2mm}p{0.2mm}p{31mm}X}
      \toprule
      \multicolumn{4}{l}{Reverse derivative trace $(\bar{z}=\prt L/\prt z)$}\\
      \multirow{9}{2mm}{\begin{tikzpicture}\draw[<-,>=triangle 60,thick](0,0)--(0,-3.8);\end{tikzpicture}} 
      & $\bar{x}_0$ & $=\bar{v}_0=\bar{v}_5\povr{v_5}{v_0}$ & $=-1.998$\\
      & $\bar{r}$ & $=\bar{v}_1=\bar{v}_3\povr{v_3}{v_1}+\bar{v}_4\povr{v_4}{v_1}$ & $=-0.01998$\\
      & $\bar{k}$ & $=\bar{v}_2=\bar{v}_4\povr{v_4}{v_2}$ & $=-0.001998$\\
      \cmidrule{2-4}
      & $\bar{v}_3$ & $=\bar{v}_5\povr{v_5}{v_3}$ & $=1.998$\\
      & $\bar{v}_4$ & $=\bar{v}_6\povr{v_6}{v_4}$ & $=-1.998$\\
      & $\bar{v}_5$ & $=\bar{v}_6\povr{v_6}{v_5}$ & $=-1.998$\\
      & $\bar{v}_6$ & $=\povr{L}{v_6}=-2(1-v_6)$ & $=-1.998$\\
      \cmidrule{2-4}
      & $\bar{L}$ & $=\povr{L}{L}$ & $=1$\\
      \bottomrule
      \end{tabularx}}
  \end{minipage}
\end{table*}

\section{Automatic differentiation}

Automatic differentiation uses the chain rule to break long calculations into small pieces, each of which can be easily differentiated \autocite{griewank08evaluating,baydin18automatic,margossian19a-review}. The chain rule tells us how to combine the differentiated pieces into the overall derivative. For example, given the function
\begin{equation*}
  f(p_1,p_2) = 2p_1p_2,
\end{equation*}
we can set $v=p_1p_2$ and obtain the partial derivative of $f$ with respect to $p_1$ by the chain rule
\begin{equation*}
  \povr{f}{p_1}=\povr{f}{v}\povr{v}{p_1}=2p_2.
\end{equation*}
The final value multiplies the two component derivatives, $\prt{f}/\prt{v}=2$ and $\prt{v}/\prt{p_1}=p_2$. To use this piece as part of a longer calculation, we substitute the numerical value of $p_2$ and store only that numerical value. For example, if $p_2=0.1$, then $\prt{f}/\prt{p_1}=0.2$

In general, long calculations can be broken into a sequence of simple steps. At each step, we calculate the numerical value of the component derivative, combine that component's value with prior component values, and then store the resulting numerical value to use in subsequent steps. Thus, we need to store only a small amount of information as we sequentially combine the component values to obtain the final overall value. The calculation has the same theoretical exactness as a complete symbolic derivative but, by substituting numerical values as we go along, we can proceed faster and with much less computational storage space. Alternatively, many optimization methods estimate the derivatives numerically, which requires relatively little storage space but has much lower accuracy. The errors in estimation impede optimization or cause optimization to fail.

The following subsections introduce two general methods of automatic differentiation and one special method for differentiating trajectories of differential equations. Those methods differ in how they apply the chain rule.

\subsection{Forward method}

The forward method begins with input parameters and then traces the chain of derivatives to the output. The center box of Table 1 illustrates the calculation for the computational graph of \Eq{euler}, shown in \Fig{graph}. That calculation yields the derivative of the output, $L$, with respect to the input parameter, $r$, written as $\dot{L}=\prt L/\prt r$.

Two early steps in the forward derivative trace show application of the chain rule. First, for $v_1=r$, we obtain $\dot{v}_1=1$, in which the overdot denotes the partial derivative with respect to $r$. Second, for $v_4$, the chain rule expands to
\begin{equation*}
  \povr{v_4}{r} = \povr{v_4}{v_2}\povr{v_2}{r} + \povr{v_4}{v_1}\povr{v_1}{r}
  	= h\lr{v_1\dot{v}_2+\dot{v}_1v_2}=0.01,
\end{equation*}
using the constant $h=0.01$ and substituting the other values given in the table. Continuing forward, we eventually arrive at the final value, $\dot{L}=-0.01998$.

The forward mode is simple and relatively easy to implement in computer code. A forward derivative trace can be done in parallel with the computations to calculate the performance value, $L$, for given inputs, as shown in the left box of Table 1.

Forward mode's primary disadvantage arises from the need to do a separate derivative trace for each input parameter. For large models with many parameters, full calculation of the gradient by forward automatic differentiation can be slow and limit application. Large optimization analyses often gain by using reverse mode automatic differentiation, which can calculate the gradient over all parameters in a single derivative trace.

\subsection{Reverse method}

The right box of Table 1 illustrates the reverse derivative trace for the same problem. Starting at the end, with $L$, and going backwards, the first step calculates $\bar{v}_6=\prt L/\prt v_6$. Then, by the chain rule, $\bar{v}_5=\prt L/\prt v_5=\bar{v}_6\,\prt v_6/\prt v_5$. By this method, the calculation continues in the reverse direction. In the final steps, at the top of that box, one can calculate $\prt L$ with respect to each input parameter, yielding the full gradient $\gL$ in one backward trace.

Modern neural network models often have millions of parameters. A single backward derivative trace calculates the full gradient of the performance function with respect to all parameters. That technique, often called back propagation in the neural network literature, provided a necessary advance in computation for the great recent breakthroughs in modern neural network modeling \autocite{goodfellow16deep}.

The same method also enhances the potential to optimize large differential equation models. However, for differential equations, one typically needs to optimize through the numerical methods used to approximate the temporal trajectories of systems. To accomplish a practical method for reverse mode differentiation through the numerical approximation algorithm required another conceptual advance \autocite{chen18neural,rackauckas20universal}.

\subsection{Differentiating trajectories of differential equations}

In essence, we apply the reverse method to the computational graph for the trajectory of the differential equation. This approach leads to a new differential equation that propagates the derivative of the performance function $L$ backward in time, similarly to how we traced the derivative of $L$ back to the parameters in the reverse mode approach of Table 1. Here, I describe the concepts in an approximate and intuitive way. I follow \textcite{chen18neural}, who provide a full explanation and complete derivation.

The goal is to calculate the derivative of $L$ as a function of the location of the trajectory at various times. For simplicity, we consider a univariate differential equation with location at a single time, $x_t$, and associated performance, $L(x_t)$. We seek the gradient of $L$ with respect to the parameters, $\bp$, and the initial condition, $x_0$.

Start with the definition $a(t) = \dd L/\dd x_t$. We then consider how this value changes with time, $t$, creating a new differential equation, $a'(t)$, with the prime denoting differentiation with respect to time
\begin{equation*}
  a'(t) = \frac{1}{h}\lr{\frac{\dd L}{\dd x_t}-\frac{\dd L}{\dd x_{t-h}}}
  		= \frac{\dd L}{h\dd x_t}\lr{1-\povr{x_t}{x_{t-h}}},
\end{equation*}
for small $h$. The term $\prt x_t/\prt x_{t-h}$ is a backward chain rule expansion, like the steps in the reverse mode trace of Table 1. Noting from \Eq{euler} that $\prt x_t/\prt x_{t-h}=1+h\prt f/\prt x_{t-h}$, and letting $h$ go to zero, we obtain
\begin{equation}\label{eq:adjoint}
  a'(t) = - a(t)\povr{f}{x_t}.
\end{equation}
As in the reverse trace of Table 1, we can trace derivatives backwards, in this case tracing back in time for $\prt x_t,\prt x_{t-h}, \ldots, \prt x_h$ to calculate the trajectory of $a(t)$. In the final backward step, we apply the derivative of $x_h$ relative to the input parameters, $\bp$, as $\prt x_h/\prt \bp$, in essence, allowing us to transform the last term of \Eq{adjoint} at time $t=h$ as
\begin{equation}\label{eq:lastStep}
  \povr{f}{x_h}\povr{x_h}{\bp}=\povr{f}{\bp}.
\end{equation}
Using that method to analyze changes in $L$ and $f$ with respect to $\bp$ instead of with respect to $x_t$, we can write the solution for \Eq{adjoint} as
\begin{equation}\label{eq:Lgrad}
  \povr{L}{\bp}=-\int_{t_1}^{t_0} a(t)\povr{f}{\bp}\dd t.
\end{equation}

Here, the backward integral trace in \Eq{Lgrad} equals the solution backward in time of the differential equation in \Eq{adjoint}, modified by a final transformation that yields the derivative of the performance function, $L$, with respect to the parameters. Standard numerical methods to analyze differential equations can be used to find the solution for \Eq{Lgrad}.

The formal derivation of \Eq{Lgrad} uses a more rigorous analysis to change $\prt f/\prt x_t$ in \Eq{adjoint} into $\prt f/\prt \bp$ in \Eq{Lgrad}, arriving at the same conclusion \autocite{chen18neural}. 

\begin{figure*}
\centering
\includegraphics[height=0.85\vsize]{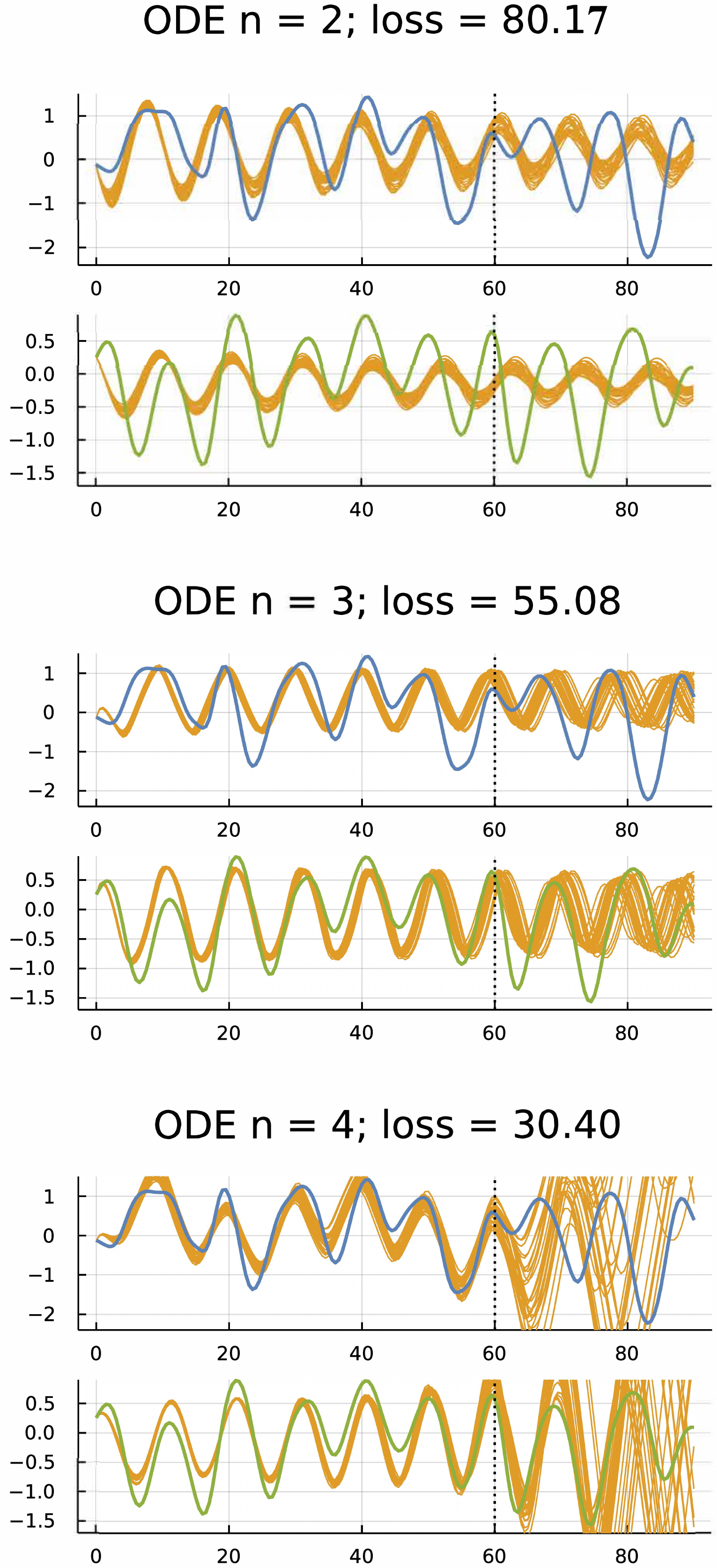}
\caption{Dynamics of hare (blue) and lynx (green) populations \autocite{odum71fundamentals,bonnaffe21neural} and fitted models (gold) \autocite{frank22optimizing}. The ordinary differential equation (ODE) models were fit to the first 61 yearly observations. The gold curves past year 60 show predictions for subsequent dynamics. For $n=2$, the model had one variable for hare and one for lynx. For $n=3,4$, the model had one or two extra variables to account for possible unobserved factors, providing more parameters to fit the data. Smaller loss means a better fit to the first 60 years. The model fitting was done by an approximate Bayesian method based on the gradient of the loss calculated by automatic differentiation \autocite{li15preconditioned}. For each model, gold trajectories arise from 30 randomly chosen parameter combinations of the Bayesian posterior distribution. From Fig.~4 of \textcite{frank22optimizing}, which provides methods, a broader analysis of these data, and other fitting approaches such as neural ODEs that use artificial neural networks to fit differential equations to target trajectories.}
\label{fig:lynx_hare}
\end{figure*}

\section{Examples}

\subsection{Fitting data for temporal trajectories}

One can fit differential equation models to time series data. For example, \Fig{lynx_hare} shows the dynamics of hare and lynx populations. The blue curve shows the fluctuation of the hare population, and the green curve shows the fluctuation of the lynx population. The gold curves show a Bayesian fit of a differential equation model to the initial 2/3 of the data and the model's prediction compared to the data for the final 1/3 of the time series.

Each model has $n$ variables, two for the log transformation of hare and lynx population sizes and $n-2$ for dummy variables tracking unobserved factors. The models seek to match the data of hare and lynx population sizes observed over 91 years in a classic ecological study summarized by \textcite{odum71fundamentals} and \textcite{bonnaffe21neural}. The differential equation for the vector of variables $\bu$ in the ODE models has the form
\begin{equation}\label{eq:odede}
  \dovr{\bu}{t} = f\lr{\bS\bu - \bb},
\end{equation}
in which the first two components of $\bu$ are the log-transformed populations sizes, the $n^2+n$ parameters are in the $n\times n$ matrix, $\bS$, and the $n$ vector, $\bb$. The function $f$ maps the $n$ dimensional input to an $n$ dimensional output, potentially inducing nonlinearity in the model. One typically selects $f$ from the set of common activation functions used in neural network models. For all runs in \textcite{frank22optimizing}, I used $f=\tanh$ applied independently to each dimension. It would be easy to study alternative ODE forms. However, the analyses in \textcite{frank22optimizing} focused on \Eq{odede}.

The three sets of plots in \Fig{lynx_hare} correspond to models with $n=2,3,4$ variables. For $n=2$, the model has one variable to track the hare population and one to track the lynx population. That model roughly matches the frequency but not the amplitude of population fluctuations during the fitting period for the initial 2/3 of the data, corresponding to the first 60 years. The model fails to predict the frequency or amplitude accurately during the prediction period years 60--90.

For $n=3$, the model adds another variable to account for unobserved factors. That model matched the fitting period better because it has an extra dummy variable and more parameters. The extra parameters determine the dynamics of the dummy variable and its interactions with the hare and lynx populations. The model's dynamics during the prediction phase roughly matches the frequency of population fluctuations. Although a bit off on amplitude, the fit is reasonably good given the very limited data and complex dynamics.

\begin{figure*}[t]
\centering
\vskip-0.3in
\begin{overpic}[width=0.99\hsize]{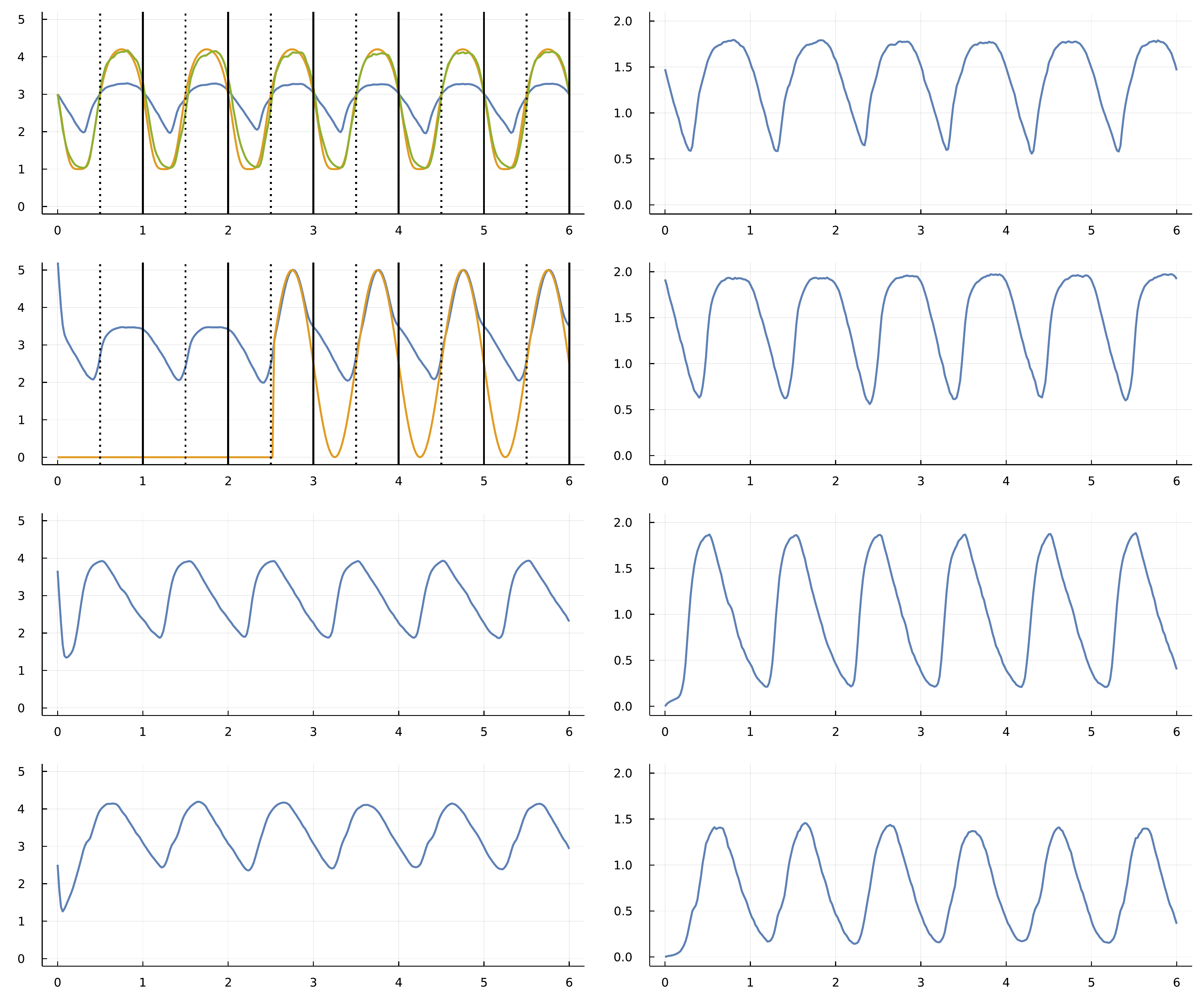}
 \put (4.5,81)    {\large (a)}
 \put (55,81)     {\large (e)}
 \put (4.5,60)    {\large (b)}
 \put (55,60)     {\large (f)}
 \put (4.5,39.3)    {\large (c)}
 \put (55,39.3)     {\large (g)}
 \put (4.5,18.4)    {\large (d)}
 \put (55,18.4)     {\large (h)}
\end{overpic}
\caption{Circadian dynamics with stochastic fluctuations and random daylight signal. (a-d) Abundance of TFs on $\log10(1+y)$ scale for $y$ molecules per cell. (e-h) Abundance of mRNAs for associated TFs. See text for explanation. From Fig.~1 of \textcite{frank22optimization}.}
\label{fig:tf}
\end{figure*}

For $n=4$, the model has two extra dummy variables and more parameters. The match to the fitting period is very good, with little variation among the sampled trajectories from the Bayesian posterior distribution of parameters. However, the match during the prediction phase is highly erratic.

Overall, the model with $n=2$ appears underfit, with insufficient parameters and flexibility to match the observed pattern or predict future observations. The model with $n=4$ appears overfit, with a close match to the fitted period but poor match for the predicted period. The model with $n=3$ appears to be a good compromise, a conclusion supported by further analyses in \textcite{frank22optimizing}. 

The Bayesian fitting procedure used automatic differentiation through a numerical differential equation solver. That approach provided a fast computational method to find a good fit to the data. The computational benefit does not alter the challenge of deciding what model to use or the criteria for deciding the relative success of different models. Instead, the method provides a simple and practical way of doing the computations that may be difficult to achieve with other methods. Technical innovation often leads to conceptual advance.

\subsection{Analyzing adaptive traits}

One can also fit a model to a theoretical challenge rather than to data. In particular, how can one build systems that produce good trajectories relative to some goal? How does natural selection design biological systems to solve environmental challenges and increase fitness?

Consider how cells control gene expression. Much of that control arises from transcription factors (TFs), which are proteins that bind to DNA and alter the rate at which nearby genes transcribe mRNA. Following the details in \textcite{frank22optimization}, we can study a simple model. 

Suppose we have $n$ genes. Each gene expresses an mRNA, which makes a TF protein. Thus, our differential equation model has to track the temporal trajectory of the numbers $n$ mRNAs within the cell and the associated $n$ TF proteins. The expression level of each gene is influenced by the abundances of the $n$ TFs. We can think of the $n$ TFs as inputs to the TF computational network, which produces $n$ outputs that influence the expression level of the $n$ genes. That input-output function depends on parameters that describe the binding of the TFs to DNA and the transformation of binding events into expression levels \autocite{bintu05transcriptional,bintu05transcriptionalb,marbach10revealing}.

We seek parameters to match a target temporal trajectory of TF abundances. Suppose we measure success by how well the first TF, labeled TF 1, follows a circadian pattern \autocite{frank22optimization}.

\Figure{tf} shows an example, in which there are $n=4$ TFs (left) and matching mRNA (right) levels. Panel (a) shows the abundance of TF 1 (blue curve) on a $\log10(1+y)$ scale for molecules per cell, $y$, over a time period of 6 days. Each day divides by the dotted vertical line, which denotes entry into daytime. The solid vertical line denotes entry into nighttime. We transform the molecular number (blue curve) into a cellular state by a commonly used Hill function \autocite{frank13input-output,zhang13ultrasensitive}, yielding the green curve that describes the cellular circadian rhythm.

We measure success by the total Euclidean distance between the gold curve that describes the environmental circadian pattern and the green curve that describes the internal cellular circadian rhythm. We search for good parameters by optimization algorithms that use the gradient of the success measure relative to the parameters of the differential equation, calculating the gradient by autodifferentiation through the numerical algorithm that calculates the temporal trajectories.

In this example, the stochastic differential equation has random perturbations of the molecular numbers per cell. This example also imposes environmental randomness, shown in panel (b). The external daylight signal (gold curve) is initially off, then comes on in the middle of day 3 and stays on for the remainder of the period shown. In general, the external signal turns on and off randomly, such that cells may pass many days entirely in the dark. The blue curve of that panel tracks the abundance of TF 2. The external daylight signal strongly stimulates production of TF 2, providing an internal signal within the cell that could be used to entrain the circadian dynamics.

The computational challenge starts with random parameters. One then searches for those parameters that lead to an internal circadian rhythm that buffers the internal stochastic molecular perturbations, uses the external daylight signal for entrainment when it is present, and maintains a good internal rhythm when the external signal is absent. The particular example illustrated in \Fig{tf} handles all of those challenges \autocite{frank22optimization}.

This model has 164 parameters. It would be difficult to find a good parameter combination without the remarkable computational advantages of automatic differentiation. In general, one can use this approach to generate and analyze hypotheses about how natural processes design biological systems to produce dynamic traits.

I finish my summary of examples with a few comments about using automatic differentiation in various applications. For the hare-lynx problem, fitting ODE models typically took several hours or less on a 2022 Apple M1 Ultra Mac Studio computer. In that study, I also fit neural ODE (NODE) models that use artificial neural networks \autocite{frank22optimizing}. Those NODE models have many more parameters and take longer to optimize but typically finish within many hours or less than one day.

For the TF problem, optimization time was typically several days, sometimes a week or more per run. These stochastic models are much more challenging problems for optimization through the differential equation solvers.

Although I optimized most individual algorithms for both problems, I made no attempt to minimize overall runtimes for any of these problems. So the times quoted here are only meant as very rough guidelines for those who might want to compare with their own approaches.

A question sometimes arises whether common derivative-free optimization methods could solve such problems as well or better than automatic differentiation. A full study of that question would require tuning the computer code to match various alternative methods. That has not been done for these problems. However, to make a quick test, I compared runs of the TF problem using automatic differentiation and the commonly used Nelder-Mead algorithm, which is a derivative-free optimization method. Using the same code with the alternative optimization methods, in three independent runs Nelder-Mead failed completely to track to circadian pattern. A single comparison run with automatic differentiation converged to a good solution.

\section{Discussion}

Many aspects of model interpretation depend on sensitivity \autocite{mester22differential}. How much do model predictions change with a change in a parameter? For example, if one wishes to change dynamics, altering a parameter with a strong effect on outcome would be better than altering a parameter with little effect.

Confidence in parameter estimates also relates to sensitivity. If large changes in a parameter have little effect on outcome, then estimates for that parameter will vary widely and confidence in a particular estimate is low.

In evolutionary models of biological traits, sensitivity may relate to genetic variability. For a parameter with low sensitivity, changes in that parameter have relatively little effect on performance. Such parameters are likely to accumulate much associated genetic variability. By contrast, changes in sensitive parameters strongly affect performance, suggesting relatively little genetic variability associated with such parameters.

Because sensitivity often means the change in performance with respect to a change in parameters, one often evaluates sensitivity by the derivative of performance with respect to parameters. For models with many parameters, automatic differentiation provides computational benefits \autocite{mester22differential}. In evolutionary analyses, the performance surface that defines sensitivity is the adaptive or fitness landscape \autocite{stadler02fitness,malan21a-survey}. Many conceptual aspects of evolutionary dynamics and of optimizing artificial neural networks come down to understanding how various factors alter the geometry of performance surfaces in models with large numbers of parameters \autocite{yang19wide}.

For studies of sensitivity and performance surface geometry, Bayesian posterior distributions of parameters provide a complementary approach. A narrow distribution typically means that small changes in a parameter provide much information, which also typically means that small changes strongly alter model outcome. In biology, narrow Bayesian posteriors may associate with less genetic variability than broad posteriors, because narrowness associates with large fitness effects for small changes in trait values.

The approximate Bayesian method to generate the predicted trajectories in \Fig{lynx_hare} provides estimates for the posterior distribution of parameters. That method depends on automatic differentiation to calculate the gradient of performance with respect to the parameters. Roughly speaking, the gradient gives the directional change of the parameters favored to improve performance in the same way that biological fitness favors changes in traits to enhance fitness.

Balanced against that directional change in improved performance, the method introduces random fluctuations in parameters, similar to the way that mutation causes random changes in traits. The balance between directional selection and random mutation creates a distribution of parameter values that approximates the Bayesian posterior distribution. Those opposing forces also match the notion of genetic variability maintained by a balance between selection and mutation. The similarity between mutation-selection genetics and computational Bayesian analysis hints at the broad conceptual relations between evolutionary dynamics and the study of optimizing large systems in artificial neural networks and other domains.

As computational techniques for automatic differentiation improve, many new opportunities for theoretical advances will arise in domains for which optimization provides an important tool. Such opportunities for theoretical application will be matched by opportunities for greater conceptual understanding of processes that improve performance.

\section*{Funding}

The Donald Bren Foundation, National Science Foundation grant DEB-1939423, and DoD grant W911NF2010227 support my research. 

\section*{Conflict of interest}

The author declares that the research was conducted in the absence of any commercial or financial relationships that could be construed as a potential conflict of interest.


\mybiblio	


\end{document}